\documentstyle[mprocl]{article}

\def\Journal#1#2#3#4{{#1} {\bf #2}, #3 (#4)}

\def\PRD{{\em Phys. Rev.} D}

\def\GRG{\em Gen. Rel. Grav.}

\def\be{\begin{equation}}
\def\ee{\end{equation}}
\def\bea{\begin{eqnarray}}
\def\eea{\end{eqnarray}}

\begin{document}

\title{FREE EVOLUTION OF NONLINEAR SCALAR FIELD COLLAPSE IN DOUBLE-NULL
COORDINATES}

\author{LIOR M. BURKO}

\address{Department of Physics, Technion---Israel Institute of
Technology, 32000 Haifa, Israel}




\maketitle\abstracts{
We study numerically the fully nonlinear spherically-symmetric 
collapse of a self-gravitating, minimally-coupled,
massless scalar field. Our numerical code is based
on double-null coordinates and
on free evolution of the metric functions and the scalar field.  
The numerical code is stable and second-order accurate.
We use this code to study the late-time asymptotic behavior
at fixed $r$ (outside the black hole), along the event horizon, and along
future null infinity. In all three asymptotic regions we find that, after
the decay of the quasi-normal modes, the perturbations are dominated by
inverse power-law tails. The corresponding power indices agree with the
integer values predicted by linearized theory.
We also study the case of a charged black hole nonlinearly
perturbed by a (neutral) self-gravitating scalar field, and find the same
type of behavior---i.e., quasi-normal modes followed by inverse power-law
tails, with the same indices as in the uncharged case.}
  
\section{Introduction}

Until recently, the late-time evolution of
non-spherical gravitational collapse was
investigated primarily in the context of linear theory. 
The late-time behavior of such perturbations has been
studied for three different asymptotic regions:
(a) at fixed $r$ (outside the black hole), (b) along null infinity, and
(c) along the future event horizon (EH). 
Qualitatively, the evolution of the linearized perturbations  
is similar in these three asymptotic regions: During the first stage,
the perturbations' shape depends strongly on the shape of the initial
data. This stage is followed by the stage of
quasi-normal (QN) ringing, Finally, there are also `tails',
characterized by an inverse power-law decay.

Case (a) was first studied by Price \cite{price}, and 
cases (b) and (c) by Gundlach {\em et al.} \cite{GPP1} It was found
that after the QN   
ringings die out, the perturbations in (a) 
decay according to $t^{-(2l+\mu+1)}$, where $\mu=1$ if there were an
initial static mode, and $\mu=2$ otherwise. Here, $l$
is the multipole moment of the mode in question, and $t$ is the
Schwarzschild time coordinate. Tails in (b) decay according
to $u_e^{-(l+\mu)}$, and in (c) according to
$v_e^{-(2l+\mu+1)}$, where $u_e$ and $v_e$ are the outgoing and ingoing
Eddington-Finkelstein coordinates, correspondingly. 

The numerical simulation of the fully-nonlinear
gravitational collapse of a spherical self-gravitating
scalar field was recently carried out, 
and the QN ringing and the `tails' were demonstarated for cases 
(a) \cite{GPP2,MC} and (c) \cite{MC}.

In what follows we shall briefly describe a recent numerical simulation of
the fully-nonlinear spherical collapse of a 
minimally-coupled massless scalar field. We
study all three cases (a), (b) and (c), and obtain values for the
power-law indices significantly closer to the linear analysis predictions
than all previous nonlinear simulations. Section \ref{sec2} describes our
numerical method, and Section \ref{sec3} describes our main results.
Further details are given elsewhere \cite{burko}.

\section{The numerical method}\label{sec2}
Our numerical code is based on free evolution of the dynamical equations
in double-null (DN) coordinates. The constraint equations are imposed only
on
the initial characteristic hypersurface, and just monitored during the
evolution. The main advantages and properties of the DN approach
are: The DN coordinates are very well adopted to the hyperbolic character
of the field equations, the interpretation of the causal structure of the
spacetime is trivial, and DN coordinates can be chosen such that the
metric is regular at the EH. Evolution is obtained by straightforward
marching along both coordinates. Because in the dynamical equations all
second derivatives are mixed, a second-order code can be obtained from
just two adjacent grid-points in each direction. Therefore, only three
grid points
are needed in each computational cell, and the second-order accuracy is
obtained by standard `predictor-corrector' technique. 
The inevitability of some sort of Dynamical Mesh Refinement for our code
can be demonstrated for
Schwarzschild (for any spacetime with an EH the reasonings are similar).
First, the EH must be included in the integration. If it were not, then
very quickly the integration domain will get very far from the EH, and
none of the cases (a), (b) or (c) would be susceptible of accurate
analysis. However, whenever the EH is included in the integration we face
a fundamental difficulty: As one moved along an outgoing null ray near the
EH, $r_{,u}$ grows exponentially, where $r$ is the area coordinate. Hence,
a small error in the location of the EH would quickly blow off. The
solution is to make the grid dense near the EH (to allow for an accurate
integration), but dilute far from the EH (to save computation time). This
is done by the following algorithms. {\em Point Splitting}: We check the
variation of the metric functions and the scalar field along ingoing
rays between two adjecent grid points. If the variation is greater than
some threshold value, we introduce an intermediate grid-point, at which we
evaluate the fields by interpolation. {\em Chopping}: The domain of
integration includes the EH. Consequently, (with vanishing charge) any
fixed $u_{\rm final}$ would lead to a crash into the singularity after a
finite lapse of time. We solve the problem by simply chopping ingoing rays
immediately after the apparent horizon, which we find locally. Chopping
enables the code to run forever while including the EH in the integration. 
Due to the
Point Splitting procedure, we obtain a denser grid than necessary far from
the EH. We introduce {\em Point Removal} to remove unnecessary
grid-points, by a method which essentially is the inverse of Point
Splitting. However, in any Kruskal-like gauge $g_{uv}$ grows rapidly along
ingoing rays, anf therefore Point Removal would be ineffective. We cure
the problem by introducing {\em Gauge Correction}, which changes the gauge
to a gauge in which $g_{uv}$ changes only slowly. The combination of 
Point Removal and Gauge Correction enables a very effective saving of
computation time. Our code was checked by the following methods: We
compared the results obtained with
different grid-parameters. We monitored the discrepancy in the two
constraint equations. We compared the mass function obtained from local
differentiation and from evolution of the `wave equation' for the mass
function. We numerically reproduced known exact solutions
[Schwarzschild, Reissner-Nordstr\"{o}m (RN), the homothetic Roberts
solution \cite{bur}]. All checks indicated
stability, convergence and second-order accuracy.

\section{Results for the decay of `tails'}\label{sec3}
We used our numerical code for the following two configurations. (1): 
Collapse over Minkowski, which leads to the formation of a
Schwarzschild-like black hole. (2): Collapse over a unity mass 
pre-existing charged background (a RN geometry) with various values of
the charge. We took an initially ingoing pulse of
squared-sine shape of compact support, and chose the amplitude to be high,
such that the final black hole mass was $M_f\approx 3.5$ in (1) and
$M_f\approx 3.8$ in (2). We then evolved the fields, and probed their
values
for cases (a), (b), and (c) in both configurations (1) and (2). In order
to approximate null infinity [for case (b)] we probed the fields on an
ingoing null ray at $v_e=10^6\;M_f$. We introduce the notion of the {\em
local power index} (LPI). Namely, we calculate the evolution in $p\equiv 
-v\;(\ln\Phi)_{,v}$ for case (c) [$v$ should be replaced by $t$ and $u$
for cases (a) and (b), respectively], where $\Phi$ is the scalar field.
The motivation behind the LPI is as follows: The power-law
behavior is just the leading-order term in an expansion in $v^{-1}$. For
any {\em finite} value of $v$ there will also be nonvanishing
contributions of the higher-order terms, which will cause a deviation from
the integer power index. In addition, any averaging or best-fit technique
would yield a result which depends on the interval. The LPI 
would thus be a fractional number, but would asymptotically approach the
integer value of the leading-order power index. Table \ref{tab:res}
summarizes our results. The agreement with the linearized-theory
predictions is remarkable, and better than the results of all previous
nonlinear
analyses.

\begin{table}[t]
\caption{Values for the Local Power Indices for cases (a), (b) and (c) in 
configurations (1) (uncharged case) and (2) (charged case). The value in
the brackets is the linearized theory prediction. \label{tab:res}}
\vspace{0.4cm}
\begin{center}
\begin{tabular}{|c|c|c|c|}
\hline
& & & \\
&
(a): along the EH &
(b): along null infinity & 
(c): along $r={\rm const}$
\\ \hline
Config. (1) &
$2.98\pm 0.01\;\;(3)$ &
$2.002\pm 0.003\;\;(2)$ &
$2.99\pm 0.02\;\;(3)$
\\ \hline 
Config. (2) &
$2.99\pm 0.01\;\;(3)$ & 
$1.996\pm 0.001\;\;(2)$ &
$2.99\pm 0.02\;\;(3)$ 
\\ \hline
\end{tabular}
\end{center}
\end{table}

\section*{Acknowledgments}
I am indebted to Amos Ori for many stimulating and helpful discussions.


\section*{References}
\begin{thebibliography}{99}

\bibitem{price}R. H. Price, \Journal{\PRD}{5}{2419}{1972}.
\bibitem{GPP1}  C. Gundlach, R. H. Price, and J. Pullin, 
\Journal{\PRD}{49}{883}{1994}.
\bibitem{GPP2} C. Gundlach, R. H. Price, and J. Pullin, 
\Journal{\PRD}{49}{890}{1994}.
\bibitem{MC}  R. L. Marsa and M. W. Choptuik, 
\Journal{\PRD}{54}{4929}{1996}.
\bibitem{burko}L. M. Burko and A. Ori, {\em Phys. Rev.} D (submitted) and
Report No. gr-qc/9703067 (unpublished).
\bibitem{bur} L. M. Burko, \Journal{\GRG}{29}{259}{1997} and references
cited therein.

\end{thebibliography}
\end{document}